\documentclass[letterpaper,11pt]{article}
\usepackage[letterpaper,top=1in, bottom=1in, left=1in, right=1in]{geometry}
\usepackage{t1enc,lscape}
\usepackage{subfigure}
\usepackage[greek,english]{babel}
\usepackage[latin3]{inputenc}
\usepackage[round]{natbib}
\usepackage{latexsym,exscale,amsfonts,amssymb,amsmath}
\usepackage{graphicx}
\usepackage[usenames,dvipsnames]{color}
\usepackage{afterpage}
\usepackage{longtable}
\usepackage{authblk}

\usepackage{setspace}
\begin{document}

\title{Tests of Lorentz Invariance Violation with Gamma Rays to probe Quantum Gravity}

\author[1]{N.\ Otte}
\author[2]{M.\ Errando}
\author[3]{S.\ Griffiths}
\author[3]{P.\ Kaaret}
\author[4]{H.\ Krawczynski}
\author[5]{A.\ McCann}
\author[6]{G.\ Sinnis}
\author[7]{F.\ Stecker}
\author[1]{I.\ Taboada}
\author[8]{V.\ Vasileiou}
\author[9]{B.\ Zitzer}
\affil[1]{School of Physics \& Center for Relativistic Astrophysics, Georgia Institute of Technology, 837 State Street, Atlanta GA 30332-0430}
\affil[2]{Department of Physics and Astronomy, Barnard College, Columbia University, NY 10027}
\affil[3]{Department of Physics and Astronomy, Van Allen Hall, University of Iowa, Iowa City, IA 52242}
\affil[4]{Physics Department, Washington University in St.\ Louis, 1 Brookings Dr., CB 1105, St.\ Louis, MO 63130}
\affil[5]{Kavli Institute for Cosmological Physics, University of Chicago, Chicago, IL 60637}
\affil[6]{Physics Division, Los Alamos National Laboratory, Los Alamos, NM 87545}
\affil[7]{NASA Goddard Space Flight Center and University of California at Los Angeles}
\affil[8]{Laboratoire Univers et Particules de Montpellier, Universit\'e Montpellier 2, CNRS/IN2P3,  CC 72, Place Eug\`ene Bataillon, F-34095 Montpellier Cedex 5, France}
\affil[9]{Argonne National Laboratory, 9700 S. Cass Avenue, Argonne, IL 60439}
\date{}
\maketitle

Special and general relativity extended our understanding of the concepts of space and time,
two of the most basic topics of investigation of modern physics. However, quantum theory has
shown that there is more to learn regarding these concepts. The group of Lorentz transformations is scale invariant.
However, considerations of how to combine the concepts of quantum mechanics and gravity (quantum gravity) indicate that there is a ``natural scale'' at which the physics of space-time
predicted by relativity theory breaks down and thus requires modification, or a new paradigm \citep{1998EJPh...19..313S}. This natural scale is the Planck scale, $E_{Pl}\equiv \sqrt{\hbar c^5/G} \simeq 1.2 \times 10^{19}$ GeV or $1.6\times10^{-35}$\,m.
Introducing such a constant scale in itself violates Lorentz invariance (LI) since relativity precludes any invariant length.
Thus, the search for a theory of quantum gravity has provided a strong motivation for testing LI violation (LIV).
Other motivations for testing and possibly modifying LI are the need to cut off high energy (UV) divergences in
quantum field theory \citep{2008LRR....11....5R} and the need for a consistent theory of black holes \citep{2011LRR....14....8S}.

While it is not possible to directly probe the Planck scale in the laboratory, effects are predicted that are testable, for example, with astroparticle experiments. Testable effects include energy dependent dispersion, maximum CR electron energy, maximum cosmic gamma-ray energy, polarization effects like birefringence, change of threshold energy for gamma-gamma pair production that modifies gamma-ray propagation in the intergalactic infrared background, modification of the GZK effect and UHECR spectrum, modification of GZK neutrino spectrum, neutrino decay, and modification of neutrino oscillations. An exhaustive review can be found in \cite{2005LRR.....8....5M}. In this paper we discuss energy dependent dispersion and threshold effects, two effects that are testable with gamma-ray instruments like the \textit{Fermi}-LAT, VERITAS, CTA, and HAWC.

\paragraph{Energy dependent dispersion}is the modification of the vacuum speed of light with energy dependent terms. Typically considered are a linear and a quadratic term. In the linear case it has been shown that  $\cal{CPT}$ is violated in effective field theory \citep{1997PhRvD..55.6760C,1998PhRvD..58k6002C}. Thus, if $\cal{CPT}$ is preserved and LI violated, the quadratic is expected to dominate. The modification of the dispersion relation may be direction dependent.\cite{2009PhRvD..80a5020K} explore this possibility quantitatively by systematically exploring $\cal{CPT}$ and Lorentz invariance violating extensions to the standard model. A sample of $\approx20$ LIV constraining observations at different positions in the sky could be used to place stringents constraints on the parameters of the standard model extension. Experiments like CTA and Hawc are well suited to deliver a sufficient number of constraints.

In practice energy dependent dispersion is tested by observing highly variable gamma-ray sources like gamma-ray bursts (GRBs), active galactic nuclei (AGN) and pulsars. Under the assumption that gamma rays of different energies are emitted simultaneously, one searches for an energy dependent propagation time. Astrophysical gamma-ray observations provide the most stringent tests because the propagation time differences increase with distance of the gamma-ray emitter as well as with energy of the gamma-ray. The best tests to date on the linear term probe scales just above the Planck energy and come from observations of GRBs with the \textit{Fermi}-LAT gamma-ray satellite \citep{2009Natur.462..331A}. Observations of two flaring AGN with the H.E.S.S.\ ($2\times10^{18}$\,GeV) \citep{2011APh....34..738H} and MAGIC ($3\times10^{17}$\,GeV) \citep{2009APh....31..226M} Cherenkov telescopes probe energy scales that are factors of 5 and 50 below the Planck energy, respectively. A constraint equal to $3\times10^{17}$\,GeV was obtained from the observation of pulsed 100\,GeV gamma-rays from the Crab ppulsar with VERITAS \citep{2012arXiv1208.2033N}. 

Much less constrained than the linear term is the quadratic term. The two AGN flares\\ \citep{2011APh....34..738H,2008PhLB..668..253M} provide limits of $6\times10^{10}$\,GeV and $6\times10^{11}$\,GeV, respectively. These limits are more constraining than the GRB observations with the \textit{Fermi}-LAT ($3\times10^{10}$\,GeV) \citep{2009Natur.462..331A} despite the closer distance and longer flaring times than GRBs due to the higher gamma-ray energies observed with Cherenkov telescopes. A comprehensive review of existing observations can be found in \cite{2011AdSpR..47..380B,2009AIPC.1112..187W}.

In the future it can be expected that these constraints will significantly improve with present and next generation gamma-ray instruments. For example, the observation of an AGN flare at the same redshift with VERITAS that is ten times more intense and has finer time structure than the one observed with H.E.S.S.\ could provide limits that are a factor ten more constraining and could thus directly probe the Planck scale. The planned Cherenkov Telescope Array (CTA) has a factor of ten improved sensitivity over VERITAS and thus will be able to resolve shorter time scales and higher energy gamma-rays from AGN. The lower energy threshold of CTA allows to probe AGN flares at higher redshifts. CTA will probe the quadratic term a factor of 50 beyond what is presently possible, while the linear term will be probed orders or magnitude beyond the Planck mass scale \citep{2012arXiv1208.5356D}. These estimates are made assuming AGN with similar flaring and redshift distributions like those that have been detected in the past and does not include the possibility of stronger flares and shorter timescales, which would provide even more sensitive tests.

A holy grail in ground based gamma-ray observations is the detection of the first GRB. A GRB detection in VHE will test LIV orders of magnitude better than is presently possible with the \textit{Fermi}-LAT due to the higher energy of the detected gamma rays. That GRBs are VHE emitters has been shown with the \textit{Fermi}-LAT with the detection of one 90\,GeV photon in the GRB reference frame \citep{2009Natur.462..331A}. Thus a low-redshift GRB should be detectable with ground based gamma-ray instruments. No GRB has been detected with VERITAS or any other IACT so far mainly because the instrument has to slew towards the GRB. Also the duty cycle of Cherenkov telescopes is limited which reduces the chances of catching a GRB \citep{2011ApJ...743...62A}. CTA will have a lower energy threshold of 20\,GeV and a faster response time than VERITAS and is more likely to detect a GRB. GRB detections in the VHE band are also expected with the water Cherenkov detector HAWC \citep{2012APh....35..641A} that has a larger field of view than VERITAS of $\approx$2~sr and a duty cycle of over 95\%. Estimated GRB detection rates with CTA and HAWC are uncertain because it is not known how the gamma-ray spectra of GRB extend from the \textit{Fermi}-LAT to higher energies. Extrapolating from the observations with the \textit{Fermi}-LAT, a rate as high as 1.5 GRBs/year and as low as one per decade can be expected for HAWC and CTA \citep{2011ExA....32..193A}. HAWC is mostly sensitive to the prompt phase of short GRBs and CTA to the afterglow of long GRBs. HAWC and CTA have limited energy resolution near the energy threshold of the instrument.  Thus, LIV constraints will most likely require simultaneous satellite detection to provide high statistics light curves at lower energies.

The third gamma-ray source class that can be used to study energy dependent dispersion is pulsars. The unexpected detection of > 100\,GeV pulsed emission from the Crab pulsar with VERITAS \citep{2011Sci...334...69V} has revived the possibility of using pulsars  to test for LIV \citep{1999A&A...345L..32K,2012arXiv1208.2033N}. In the next five years VERITAS will regularly observe the Crab pulsar, which allows to continuous improve the existing LIV constraint to a level competitive with the best available limits from ground based observations due to increasing the exposure, reaching out to higher energies, and improving the analysis. So far only one pulsar has been detected in the VHE band. By detecting new pulsars in VHE with VERITAS - in particular millisecond pulsars - LIV limits ten times more sensitive than the present Crab pulsar limit could be provided. CTA with its lower energy threshold of 20\,GeV will have direct access to the tail of the pulsed emission from the $\approx150$ detected gamma-ray pulsars with the \textit{Fermi}-LAT and will thus provide a unique opportunity to sensitively test LIV effects in all directions at the Planck scale.  

Tests of LIV with individual objects are ultimately limited by source intrinsic effects that could either hide or fake an energy dependent dispersion. In some observations this is already the case, e.g.\ the observation of a shift in an AGN flare \citep{2008PhLB..668..253M,2007ApJ...669..862A} or the low statistics at high energies paired with a rich structure at lower energies as seen in GRBs \citep{2009Natur.462..331A}. These limitations can be overcome by a) a better understanding of the source engine, b) testing with as many different source classes as possible, and c) using sources at as many different distances as possible. The last of these allows unambiguous separation of source intrinsic and propagation effects, although in the case of pulsars where an LIV-like effect is observed a long-term observation program can also differentiate between a source intrinsic and a propagation effect \citep{2012arXiv1208.2033N}. Gamma-ray instruments like CTA and HAWC will be key in these studies. 

\paragraph{Threshold effects} arise if the maximum possible speed of electrons deviates from the vacuum speed of light and are discussed in more detail in \cite{2001APh....16...97S}. Two different scenarios are possible. In the subluminal case the maximum speed is less than the vacuum speed of light and in superluminal case the maximum speed is larger. In the subluminal case it would be kinematically  allowed for gamma-ray photons to decay into electron positron pairs. From observations of 50\,TeV gamma-rays from the Crab Nebula it can be deduced that the maximum speed of electrons deviates by less than $2\times10^{-16}$ from the vacuum speed of light. In the superluminal case electrons would be able to radiate vacuum Cherenkov radiation thus limiting the maximum energies of electrons. Direct observations of electrons in the cosmic ray spectrum are not very constraining due to a maximum observed energy in the TeV range. Another consequence of superluminal motion is that the threshold for pair production is increased which would reduce the gamma-ray opacity from extragalactic background light absorption \citep{1999ApJ...518L..21K,2001APh....16...97S}. The observation of 20\,TeV gamma-rays from the AGN Mkn 501 \citep{2001A&A...366...62A} constrains that the speed of electrons can not be larger than $1+1.3\times10^{-15}$ the speed of light. In the future only modest observational improvements (factor five) can be expected on these constrains with ground based experiments as this would require the observation of gamma rays at much higher energies than are observable in the foreseeable future.    

\bibliographystyle{unsrtnat}
\bibliography{bibliography}
\end{document}